\documentclass{article}
\usepackage{amssymb}

\input{tcilatex}

\begin{document}

\title{Photon-assisted electron transmission resonance through a quantum
well with spin-orbit coupling}
\author{Cun-Xi Zhang$^1$, Y.-H. Nie$^{1,2,}$\thanks{
E-mail: nieyh@sxu.edu.cn} and J.-Q. Liang$^1$ \\
%EndAName
$^1${\small Institute of Theoretical Physics and Department of Physics,}\\
{\small \ Shanxi University, Taiyuan, Shanxi 030006, China}\\
$^2${\small Department of Physics, Yanbei Normal Institute, } {\small %
Datong, Shanxi 037000, China}}
\maketitle

\begin{abstract}
Using the effective-mass approximation and Floquet theory, we study the
electron transmission over a quantum well in semiconductor heterostructures
with Dresselhaus spin-orbit coupling and an applied oscillation field. It is
demonstrated by the numerical evaluations that Dresselhaus spin-orbit
coupling eliminates the spin degeneracy and leads to the splitting of
asymmetric Fano-type resonance peaks in the conductivity. In turn, the
splitting of Fano-type resonance induces the spin- polarization-dependent
electron-current. The location and line shape of Fano-type resonance can be
controlled by adjusting the oscillation frequency and the amplitude of
external field as well. These interesting features may be a very useful
basis for devising tunable spin filters.
\end{abstract}


The Fano-type resonance, which arises from interference between a localized
state and the continuum band\cite{Fano}, has been widely known across many
different branches of physics and observed in a great variety of experiments
including atomic photoionization\cite{Fano(1965)}, electron and neutron
scattering\cite{Simpson}, Raman scattering\cite{Cerdeira}, and
photoabsorption in quantum well structures\cite{Feist,Schmidt}. In recent
years, Fano-type resonances have been reported to appear also in the
experiments on electron transport through mesoscopic systems. Fano-type
resonances in electronic transport through a single-electron transistor
allow one to alter the interference between the two paths by changing the
voltages on various gates\cite{Gores}. Reference\cite{Kobayashi} reported
the first tunable Fano experiment in which a well-defined Fano system is
realized in an Aharonov-Bohm(AB) ring with a quantum dot(QD) embedded in one
of its arms, which is the first convincing demonstration of this effect in
mesoscopic systems. The Fano effect in a quantum wire with a side-coupled
quantum dot occurs in a way different from that in QD-AB-ring system,
because only reflected electrons at the QD are involved for its emergence%
\cite{Franco,Kobayashi2,Sato}. However, except Ref.\cite{Shelykh} which
reported the first Fano-type resonances due to the interaction of the
electron states with opposite spin-orientation, so far the Fano effects have
been studied theoretically and experimentally for electron transport through
mesoscopic systems but not considering, to our knowledge, the electron spin
degree of freedom which plays an important role in spintronics. Recently, a
large number of spin-dependent phenomena---for instance, Datta-Das spin
field-effect transistor\cite{Datta} spin transport\cite{McGurie}, spin Hall
effect\cite{Hirsch}, spin-dependent tunneling phenomena in semiconductor
heterostructures\cite{Yassievich}, and so on---have attracted great
attention because of the prospect of technological applications.

In general, the condition for the Fano-type resonance to occur is the
presence of two scattering channels at least: the discrete level and
continuum band. In the mesoscopic systems mentioned above, the nature of two
scattering channels is dependent on the geometry of the device under
consideration. In this paper we adopt the interferometer geometry which is
realized by a time-periodic quantum well in semiconductor heterostructures
and study the spin-dependent Fano resonance and electron transport over
through the quantum well with Dresselhaus spin-orbit coupling. Floquet
scattering through a time-periodic potential induces mixing of the continuum
and bound states and leads to the appearance of asymmetric Fano-type
resonances in the conductivity. The effect of Dresselhaus spin-orbit
coupling on the motion of electrons makes the effective band mass of
electrons dependent on the spin-orientation; consequently, the asymmetric
Fano-type resonance splits into two different peaks corresponding to
electrons with opposite spin polarization, which leads to the
spin-polarization-dependent transmission electron-current.

We consider the transmission of a single electron with incident wave vector $%
k=\left( k_{||},k_{z}\right) $ through a one-dimension time-periodic
potential well which extends from $0$ to $a$---that is,

\begin{equation}
V(z,t)=\QATOPD\{ . {0,\text{ \ \qquad \qquad ~\qquad \qquad }z<0\text{ and }%
z>a}{-V_{0}+V_{1}\cos (\omega t),\text{ \qquad \quad \quad }0\leqslant
z\leqslant a},
\end{equation}%
here the well depth $V_{0}$ grows along $z$ $\Vert $ $\left[ 001\right] $
and can be adjusted by an applied field, $k_{||}$ is the wave vector
parallel to the plane of the well, and $k_{z}$ is the wave vector normal to
the well plane along the direction of tunnelling. The electron passes three
regions from the left to the right with $I$ , $III$ and $II$ denoting the
right, left and central regions, respectively. We assume a low enough
temperature such that the electron-phonon interactions considered in Ref.%
\cite{M.Lax} can be neglected. Thus the electron motion in semiconductor
heterostructures may be described by a time-dependent Schr\"{o}dinger
equation 
\begin{equation}
i\hbar \frac{\partial }{\partial t}\Phi (z,s,t)=\hat{H}\Phi (z,s,t)
\end{equation}%
with Hamiltonian

\begin{equation}
\hat{H}=-\frac{\hbar ^{2}}{2\mu }\frac{\partial ^{2}}{\partial z^{2}}+\frac{%
\hbar ^{2}k_{||}^{2}}{2\mu }+V(z,t)+\hat{H}_{D},
\end{equation}%
\begin{equation}
\hat{H}_{D}=\left\{ \QATOP{0,\qquad \qquad \quad \qquad \quad \qquad \quad
\quad \quad \quad \quad \quad \quad \quad \quad z<0\text{ and }z>a}{\gamma %
\left[ \hat{\sigma}_{x}k_{x}\left( k_{y}^{2}-k_{z}^{2}\right) +\hat{\sigma}%
_{y}k_{y}\left( k_{z}^{2}-k_{x}^{2}\right) +\hat{\sigma}_{z}k_{z}\left(
k_{x}^{2}-k_{y}^{2}\right) \right] ,\quad 0\leqslant z\leqslant a}\right. .
\end{equation}%
Where $\mu $ is the effective mass of electron and $\hat{H}_{D}$ denotes the
spin-dependent Dresselhaus term in Zinc-blende structure semiconductors\cite%
{Dresselhaus}. $\hat{\sigma}_{\alpha }$ is the Pauli matrices, and $\gamma $
a material constant describing the strength of the spin-orbit coupling. We
have assumed that $\gamma \approx 0$ outside the well. In the quantum well
with finite depth the confinement of the electron wave function in the
growth direction forces quantization of the corresponding component of the
wave vector; thus, one should consider $k_{z}$ in Hamiltonian as an operator 
$-i\partial /\partial z$. Assuming that the kinetic energy of the incidence
electron is much smaller than the well depth $V_{0}$ the Dresselhaus term $%
\hat{H}_{D}$ may be simplified to\cite{Yassievich} 
\begin{equation}
\hat{H}_{D}=\left\{ \QATOP{0,\qquad \qquad \quad \qquad \quad \quad z<0\text{
and }z>a}{\gamma \left( \hat{\sigma}_{x}k_{x}-\hat{\sigma}_{y}k_{y}\right) 
\frac{\partial ^{2}}{\partial z^{2}},\quad 0\leqslant z\leqslant a\qquad }%
\right. .
\end{equation}%
Using Pauli matrices the Hamiltonian (3) inside potential well can be
written as 
\begin{equation}
\hat{H}=\left[ 
\begin{array}{rr}
-\frac{\hbar ^{2}}{2\mu }\frac{\partial ^{2}}{\partial z^{2}}+\frac{\hbar
^{2}k_{||}^{2}}{2\mu }+V(z,t) & \gamma k^{+}\frac{\partial ^{2}}{\partial
z^{2}}\quad \quad \\ 
\gamma k^{-}\frac{\partial ^{2}}{\partial z^{2}}\quad \quad & -\frac{\hbar
^{2}}{2\mu }\frac{\partial ^{2}}{\partial z^{2}}+\frac{\hbar ^{2}k_{||}^{2}}{%
2\mu }+V(z,t)%
\end{array}%
\right]
\end{equation}%
with $k^{\pm }=k_{x}\pm ik_{y}$. Inserting $\Phi (z,s,t)=\psi (z,t)\chi _{s}$
with $\chi _{s}$ the spin wave function and Eq.(6) into Schr\"{o}dinger
Eq.(2) we obtain

\begin{equation}
\chi _{\pm }=\frac 1{\sqrt{2}}\binom 1{\mp e^{-i\varphi }}
\end{equation}
which describes the electron spin states of the opposite spin polarizations.
Here $\varphi $ is the polar angle of the wave vector $\vec{k} $ in the $xy$
plane, $\vec{k}_{\Vert }=\left( k_{\Vert }\cos \varphi ,k_{\parallel }\sin
\varphi \right) $, The orientation of electron spins $\vec{s}_{\pm }(\vec{k}%
_{||})=\chi _{\pm }^{+}\hat{\sigma}\chi _{\pm }=\left( \mp \cos \varphi ,\pm
\sin \varphi ,0\right) $ corresponding to the eigenstates $\chi _{\pm }$
depends on the direction of the wave vector $\vec{k}_{||\text{ }}$in the $xy$
plane. In the spin subspace the Hamiltonian $\hat{H}$ may be reduced as

\begin{equation}
\hat{H}_{\pm }=\chi _{\pm }^{+}\hat{H}\chi _{\pm }=-\frac{\hbar ^{2}}{2\mu
_{\pm }}\frac{\partial ^{2}}{\partial z^{2}}+\frac{\hbar ^{2}k_{\Vert }^{2}}{%
2\mu }+V(z,t),
\end{equation}%
where the modified effective mass of the electron depends not only on
Dresselhauss coupling constant $\gamma $ and in-plane electron wave vector k$%
_{||\text{ }},$ but also on the orientation of the electron spin, and is
given by $\mu _{\pm }=\mu \left( 1\pm \gamma 2\mu k_{\Vert }/\hbar
^{2}\right) ^{-1}$. Although the modification of the electron effective mass
is very small, it plays an important role in the generation of the
asymmetric Fano resonance peak-splitting. Thus the Schr\"{o}dinger equation
can be rewritten as

\begin{equation}
i\hbar \frac \partial {\partial t}\Phi _{\pm }(z,t)=\hat{H}_{\pm }\Phi _{\pm
}(z,t)
\end{equation}
with 
\begin{equation}
\Phi _{\pm }(z,t)=\chi _{\pm }\psi _{\pm }(z,t)\exp (i\vec{k}_{||}\cdot \vec{%
\rho})
\end{equation}
where $\vec{\rho}$ $=(x,y)$ is a vector in the well-plane. Inserting Eq.(8)
and (10) into Eq.(9) we obtain the equation of $\psi _{\pm }(z,t)$

\begin{equation}
i\hbar \frac{\partial }{\partial t}\psi _{\pm }(z,t)=-\frac{\hbar ^{2}}{2\mu
_{\pm }}\frac{\partial ^{2}}{\partial z^{2}}\psi _{\pm }(z,t)+\frac{\hbar
^{2}k_{\Vert }^{2}}{2\mu }\psi _{\pm }(z,t)+V(z,t)\psi _{\pm }(z,t).
\end{equation}%
Inside the potential well, the potential $V(z,t)$ in Eq.(11) is a periodic
function of time, thus, according to the Floquet theorem\cite%
{Shirley,Holthaus,Fromherz} Eq.(11) has a solution of the form

\begin{equation}
\psi _{\pm }^F(z,t)=\varphi _{\pm }(z,t)\exp (-iE_F^{\pm }t/\hbar ),
\end{equation}
where $E_F^{\pm }$ is the Floquet energy eigenvalue and $\varphi _{\pm
}(z,t) $ is a periodic function of time: $\varphi _{\pm }(z,t)=\varphi _{\pm
}(z,t+T)$ with period $T=\frac{2\pi }\omega .$ Substituding Eq.(12) into
(11) for $\varphi _{\pm }(z,t)=g_{\pm }(z)f_{\pm }(t),$ we have two
separated equations with an introduced constant $E^{\pm }$\cite%
{Burmeister,Li}

\begin{equation}
-\frac{\hbar ^2}{2\mu _{\pm }}\frac{d^2}{dz^2}g_{\pm }(z)=\left( E^{\pm
}+V_0-\frac{\hbar ^2k_{\Vert }^2}{2\mu }\right) g_{\pm }(z),
\end{equation}

\begin{equation}
i\hbar \frac d{dt}f_{\pm }(t)-V_1\cos (\omega t)f_{\pm }(t)=\left( E^{\pm
}-E_F^{\pm }\right) f_{\pm }(t).
\end{equation}
The solution of Eq.(14) is found as

\begin{equation}
f_{\pm }(z,t)=\exp [-i(E^{\pm }-E_F^{\pm })t/\hbar ]\sum_{n=-\infty
}^{+\infty }J_n\left( \frac{V_1}{\hbar \omega }\right) \exp \left( -in\omega
t\right) .
\end{equation}
Where we have taken the initial condition $f_{\pm }(0)=1$, and $J_n(x)$ is
the n-th order Bessel function of the first kind. Since $f_{\pm }(t)=f_{\pm
}(t+T)$, Eq. (14) requires that $E^{\pm }=E_m^{\pm }=E_F^{\pm }+m\hbar
\omega $ with $m$ being an integer.

The incoming and outgoing waves (channels) of Floquet scattering form the
sidebands (or Floquet channels) with energy spacing $\hbar \omega $
according to $E_{m}^{\pm }=E_{F}^{\pm }+m\hbar \omega $ ($m$ is the sideband
index). The mode of $E_{m}<0$ is an evanescent mode, and the corresponding
sideband is called an evanescent sideband because such a mode with imaginary 
$k_{m}$ can not propagate\cite{Li,Bagwell}. The equation of $g_{\pm }(z)$
has a solution

\begin{equation}
g_{\pm }(z)=\sum_{m=-\infty }^{+\infty }[a_m^{\pm }e^{iq_m^{\pm }z}+b_m^{\pm
}e^{-iq_m^{\pm }z}],
\end{equation}
where $a_m^{\pm }$ and $b_m^{\pm }$ are constant coefficients and $q_m^{\pm
}=\left[ \frac{2\mu _{\pm }}{\hbar ^2}(E_F^{\pm }+m\hbar \omega -\frac{\hbar
^2k_{||}^2}{2\mu }+V_0)\right] ^{1/2}$. Thus the wave function inside the
oscillation-potential well can be expressed as

\begin{eqnarray}
\Phi _{\pm }^{II}(z,t) &=&\chi _{\pm }\sum_{n=-\infty }^{+\infty
}\sum_{m=-\infty }^{+\infty }[a_m^{\pm }e^{iq_m^{\pm }z}+b_m^{\pm
}e^{-iq_m^{\pm }z}]J_{n-m}(\frac{V_1}{\hbar \omega })e^{-iE_{zn}^{\pm
}t/\hbar }  \nonumber \\
&&\cdot \exp (i\vec{k}_{||}\cdot \vec{\rho}-iE_{||}^{\pm }t/\hbar ).
\end{eqnarray}
Since electrons incident to the oscillating region will be scattered
inelastically into an infinite number of Floquet sidebands, so the wave
function outside the well can be written as the superposition of waves with
all values of energy:

\begin{equation}
\Phi _{\pm }^I(z,t)=\chi _{\pm }[e^{ik_{z0}^{\pm }z-iE_{z0}^{\pm }t/\hbar
}+\sum_{n=-\infty }^{+\infty }r_{n0}^{\pm }e^{-ik_{zn}^{\pm }z-iE_{zn}^{\pm
}t/\hbar }]\exp (i\vec{k}_{||}\cdot \vec{\rho}-iE_{||}^{\pm }t/\hbar ),
\end{equation}

\begin{equation}
\Phi _{\pm }^{III}(z,t)=\chi _{\pm }\sum_{n=-\infty }^{+\infty }t_{n0}^{\pm
}e^{ik_{zn}^{\pm }z-iE_{zn}^{\pm }t/\hbar }\exp (i\vec{k}_{||}\cdot \vec{\rho%
}-iE_{||}^{\pm }t/\hbar ),
\end{equation}%
where $E_{z0}^{\pm }+E_{\Vert }^{\pm }=E_{0}^{\pm },$ $E_{\Vert }^{\pm }=%
\frac{\hbar ^{2}k_{\Vert }^{2}}{2\mu _{1}}$ and $k_{zn}^{\pm }=\sqrt{\frac{%
2\mu _{1}}{\hbar ^{2}}(E_{z0}^{\pm }+n\hbar \omega )}.$ For the sake of
simplicity we consider the case that $E_{0}^{\pm }<\hbar \omega $
corresponding to the propagating mode of the lowest energy. $r_{n0}^{\pm }$
and $t_{n0}^{\pm }$ are the probability amplitudes of the reflecting waves
and outgoing waves from the sideband $0$ to sideband $n$, respectively. The
continuity of $\Phi _{\pm }$ and $\frac{1}{\mu }\frac{\partial }{\partial z}%
\Phi _{\pm }$ at the interfaces $z=0$ and $z=a$ requires

\begin{eqnarray}
\sum_{m=-\infty }^{+\infty }J_{n-m}(\frac{V_{1}}{\hbar \omega })(a_{m}^{\pm
}+b_{m}^{\pm }) &=&\delta _{n0}+r_{n0}^{\pm }, \\
\frac{\mu _{1}}{\mu _{\pm }}\sum_{m=-\infty }^{+\infty }J_{n-m}(\frac{V_{1}}{%
\hbar \omega })q_{m}^{\pm }(a_{m}^{\pm }-b_{m}^{\pm }) &=&k_{zn}^{\pm
}(\delta _{n0}-r_{n0}^{\pm }),
\end{eqnarray}%
\begin{eqnarray}
\sum_{m=-\infty }^{+\infty }J_{n-m}(\frac{V_{1}}{\hbar \omega }%
)(e^{iq_{m}^{\pm }a}a_{m}^{\pm }+e^{-iq_{m}^{\pm }a}b_{m}^{\pm })
&=&e^{ik_{zn}^{\pm }a}t_{n0}^{\pm }, \\
\frac{\mu _{1}}{\mu _{\pm }}\sum_{m=-\infty }^{+\infty }J_{n-m}(\frac{V_{1}}{%
\hbar \omega })q_{m}^{\pm }(e^{iq_{m}^{\pm }a}a_{m}^{\pm }-e^{-iq_{m}^{\pm
}a}b_{m}^{\pm }) &=&k_{zn}^{\pm }e^{ik_{zn}^{\pm }a}t_{n0}^{\pm }.
\end{eqnarray}%
The continuity conditions of the wave functions Eq.(20 ) to (23 ) can be
expressed as the matrix forms

\begin{equation}
\left\{ 
\begin{array}{c}
J(A+B)=\triangle +R \\ 
\frac{\mu _{1}}{\mu _{\pm }}JQ(A-B)=K(\triangle -R) \\ 
J(LA+L^{-1}B)=ST \\ 
\frac{\mu _{1}}{\mu _{\pm }}JQ(LA-L^{-1}B)=KST%
\end{array}%
\right.
\end{equation}%
with the help of the square matrixes defined by the matrix elements $%
J_{nm}=J_{n-m}(\frac{V_{1}}{\hbar \omega }),$ $Q_{nm}=q_{n}^{\pm }\delta
_{nm},$ $K_{nm}=k_{n}^{\pm }\delta _{nm},$ $S_{nm}=e^{ik_{zn}^{\pm }a}\delta
_{nm},$ and $L_{nm}=e^{iq_{m}^{\pm }a}\delta _{nm}$ and the column matrixes
by the matrix elements $A_{n}=a_{n}^{\pm },\,B_{n}=b_{n}^{\pm },\,\Delta
_{n}=\delta _{n0},\,R_{n}=r_{n0}^{\pm },\,$and $T_{n}=t_{n0}^{\pm }.$ Where $%
R$ and $T$ denote the matrices of reflecting and transmission amplitudes.
From the matrix equation (24 ) one can obtain the matrix of the transmission
amplitude

\begin{equation}
T=S^{-1}(M_{2}^{-1}L^{-1}M_{1}-M_{1}^{-1}LM_{2})(M_{2}^{-1}M_{1}-M_{1}^{-1}M_{2})\triangle ,
\end{equation}%
where $M_{1}=[J^{-1}+\frac{\mu _{\pm }}{\mu _{1}}Q^{-1}J^{-1}K]$ and $%
M_{2}=(J^{-1}-\frac{\mu _{\pm }}{\mu _{1}}Q^{-1}J^{-1}K)$ . The total
electron-transmission probabilities of spin-up and spin-down components are
given by

\begin{equation}
T^{\pm }=\sum_{m=0}^{+\infty }\left| t_{m0}^{\pm }\right| ^2
\end{equation}
from which the conductance of the electrons through semiconductor
heterostructures can be obtained by Landauer-Buttiker formula\cite%
{Landauer,Buttiker}

\begin{equation}
G^{\pm }=\frac{2e^2}hT^{\pm }=\frac{2e^2}h\sum_{m=0}^{+\infty }\left|
t_{m0}^{\pm }\right| ^2.
\end{equation}

We now study numerically the scattering of a incident wave by an oscillating
quantum well of InP-GaSb-InP semiconductor hereostructures and calculate the
conductivity with the help of Eq.(25)--(27). The minimum number of sidebands
needed to be included in the sum of Eq.(26) depends on the oscillation
amplitude of the potential well. In general, it is enough to take $%
N>V_{1}/\hbar \omega $\cite{Bulgakov}.

The mechanism of resonance considered here is different from Ref.\cite{M.Lax}
where the resonance originates from an accumulation of electrons in bound
states of the well (this accumulation of electrons conversely produces
strong feedback on the transmission in the incident channel). In our model,
the interaction of electrons with the oscillating field leads to
photon-mediated transmission resonances. The incident electrons can emit
photons and drop to the bound states of the potential well. Similarly, the
electrons in bound states can also jump to incident channels or other
Floquet channels by absorbing photons. This forms the discrete channel of
scattering required by Fano-type resonance. Once the energy difference
between the incident electrons and the bound states of the well is equal to
integer times the energy of one photon, transmission resonance occurs. In
Fig.(1), we plot the conductivity $G^{\pm }$ as a function of the incident
electron energy $E_{z0}$ for $V_{0}=300meV,\,V_{1}=10meV,\,\,k_{\Vert
}=10^{6}cm^{-1},\,\hslash \omega =10meV.$ The parameters of the
semiconductor heterostructure are chosen as $\,\gamma =187eV\cdot $ \AA $%
^{3},\,\mu =0.041m_{e}$ ( $m_{e}$ the mass of the free electron) for GaSb, $%
\gamma _{1}=8eV\cdot $\AA $^{3},\,\mu _{1}=0.081m_{e}$ for InP according to
Ref.\cite{Yassievich}. In actual calculation we take $\gamma _{1}\approx 0$
because $\gamma >>\gamma _{1}.$ Sidebands of $n=0,\,\pm 1,\,\cdots ,\,\pm 5$
are taken into account so that $T^{\pm }=\sum_{n=0}^{5}|t_{n0}^{\pm }|^{2}.$
The conductivity pattern in Fig.(1) shows two obvious asymmetric resonance
peeks, at $E_{z}^{+}=5.2\,meV$ for the spin-up state \textquotedblleft
+\textquotedblright\ and $E_{z}^{-}=2.44\,meV$ for the spin-down state
\textquotedblleft -\textquotedblright . The corresponding bound state
energies are $E_{b}^{+}=-4.796\,meV$ , $E_{b}^{-}=-7.559$ $meV$ for $V_{1}=0$
, and resonance energies satisfy the relation $E_{z}^{\pm }=\hslash \omega
+E_{b}^{\pm }.$ When $E_{z}$ increases continuously and enters the second
incident channel---$i.e.,$ $\hslash \omega <E_{z}<2\hslash \omega $---the
second set of asymmetric resonance peeks appears(see the inset of Fig.(1)),
but they are very small because the probability of a two-photon process is
much less than that of one-photon process. For an incident electron of given
energy there exist similar resonance peaks in the conductivity when the
oscillation frequency increases.

The amplitude of oscillating field indicates actually the strength of
electron coupling with the external field. The correspondence between the
resonance location and energy level of the potential well makes sense only
when, strictly speaking, $V_{1}\rightarrow 0$. The electron coupled with the
applied field leads to the broadening of the level and thus the asymmetric
resonance can take place within certain range of energy rather than at a
single level. Figure 2 shows the variance of spectrum in which the
asymmetric resonance peaks are getting \textquotedblleft\ fat
\textquotedblright\ with increasing oscillation amplitude for spin states
\textquotedblleft +\textquotedblright\ and \textquotedblleft
-\textquotedblright , which provides a means to adjust the line shape of
Fano-type resonance by external parameters. When the amplitude of the
oscillating field is very small, the effect of energy level broadening is
inconspicuous and the Fano-type resonance displays sharp peaks. Thus the
bound-state energy can be determined by location of the sharp resonance
peaks according to $E_{z}^{\pm }=\hslash \omega +E_{b}^{\pm },$ which can be
used to measure the structure parameters of semiconductor herostructures
such as thickness and band gaps and even spin-orbit coupling constant.

The locations of the asymmetric resonance peaks in energy parameter space
depend not only on the width and depth of the well, but also on the
oscillation frequency of external field. In Fig. 3, we plot the frequency
dependence of the conductivity for spin states \textquotedblleft
+\textquotedblright\ and \textquotedblleft -\textquotedblright . The
asymmetric resonance peaks move toward the direction of high energy as the
frequency increases. It is apparent that the location control of resonance
peaks is easien by adjusting the field frequency than the parameters of the
semiconductor heterostructure in practical experiments.

The spin-polarization-dependent splitting of Fano-type resonance peaks is
advantageous to the realization of spin current. Figure 4 shows the
spin-polarization-dependent transmission of electrons as a function of the
electron energy according to $P=(T^{+}-T^{-})/(T^{+}+T^{-})$. Figure 4
indicates that the transmission current with the spin \textquotedblleft
-\textquotedblright\ polarization (solid line) is dominate [over 80\%
comparing with spin\textquotedblleft +\textquotedblright\ polarization
current (dashed line)] for the energy range $E_{z}$ from $1\,meV$ to $%
3.5\,meV$ $.$ The characteristics of the energy $E_{z}$ dependence of the
spin-polarization current suggests that one can exploit the Fano-type
resonances as the basis of a spin filter and control the external
parameters---e.g., the oscillation frequency $\omega $ and the amplitude $%
V_{1}$ of the external field---to tune the energy to align with specific
resonances.

In summary, we have investigated theoretically the property of the electron
transport through a quantum well in semiconductor heterostructures with
spin-orbit coupling and an applied field. The numerical results demonstrate
that Dresselhaus spin-orbit coupling eliminates the spin degeneracy and
leads to a splitting of asymmetric Fano-type resonance peaks in the
conductivity. The spin polarization arising from the splitting of Fano-type
resonance has an advantage to realize the spin current. The location and
line shape of Fano-type resonance can be controlled by adjusting the
oscillation frequency $\omega $ and amplitude $V_{1}$ of the external field.
These interesting features not only deepen our fundamental understanding of
the role of spin-orbit coupling in solids, but also may be a very useful
basis for devising tunable spin filters.

This work was supported by the National Nature Science Foundation of China
(Grant No. 10475053) and Shanxi Nature Science Foundation (Grant No.
20051002).

Figure captions

Fig.1 Splitting of Fano-type resonance in Conductivity $G^{\pm }$ for $%
V_{0}=300meV,\,V_{1}=10meV,\,\,a=60$\AA $,$ $k_{\Vert
}=10^{6}cm^{-1},\,\hslash \omega =10\,meV,\,\,\mu =0.041m_{e},$ $\mu
_{1}=0.081m_{e},$ $\gamma =187eV\cdot $\AA $^{3},$ and $\gamma _{1}=0.$ The
inset is a detail of the tiny resonance peaks resulting from electrons
exchanging two photons with applied field.

\bigskip

Fig.2 Dependance of the width of Fano-type resonance peaks in conductivity $%
G^{\pm }$ on amplitude of applied field $V_{1}$ for spin states
\textquotedblleft +\textquotedblright\ and \textquotedblleft
-\textquotedblright\ with the parameters in Fig. 1.

\bigskip

Fig.3 Conductivity $G^{\pm }$ as a function of $E_{z}$ for different $%
\hslash \omega ,$ indicating that resonance peaks move toward the direction
of high energy as $\hslash \omega $ increases, with $V_{1}=15\,meV$ and the
other parameters in Fig. 1.

\bigskip

Fig.4 Dependence of the polarization efficiency of the transmission
electrons on the energy of incident electron and spin orientation. The solid
curve denotes polarization along spin \textquotedblleft -\textquotedblright\
and the dashed curve denotes spin \textquotedblleft +\textquotedblright\
with $\,V_{1}=30\,meV$ and the other parameters in Fig. 1.

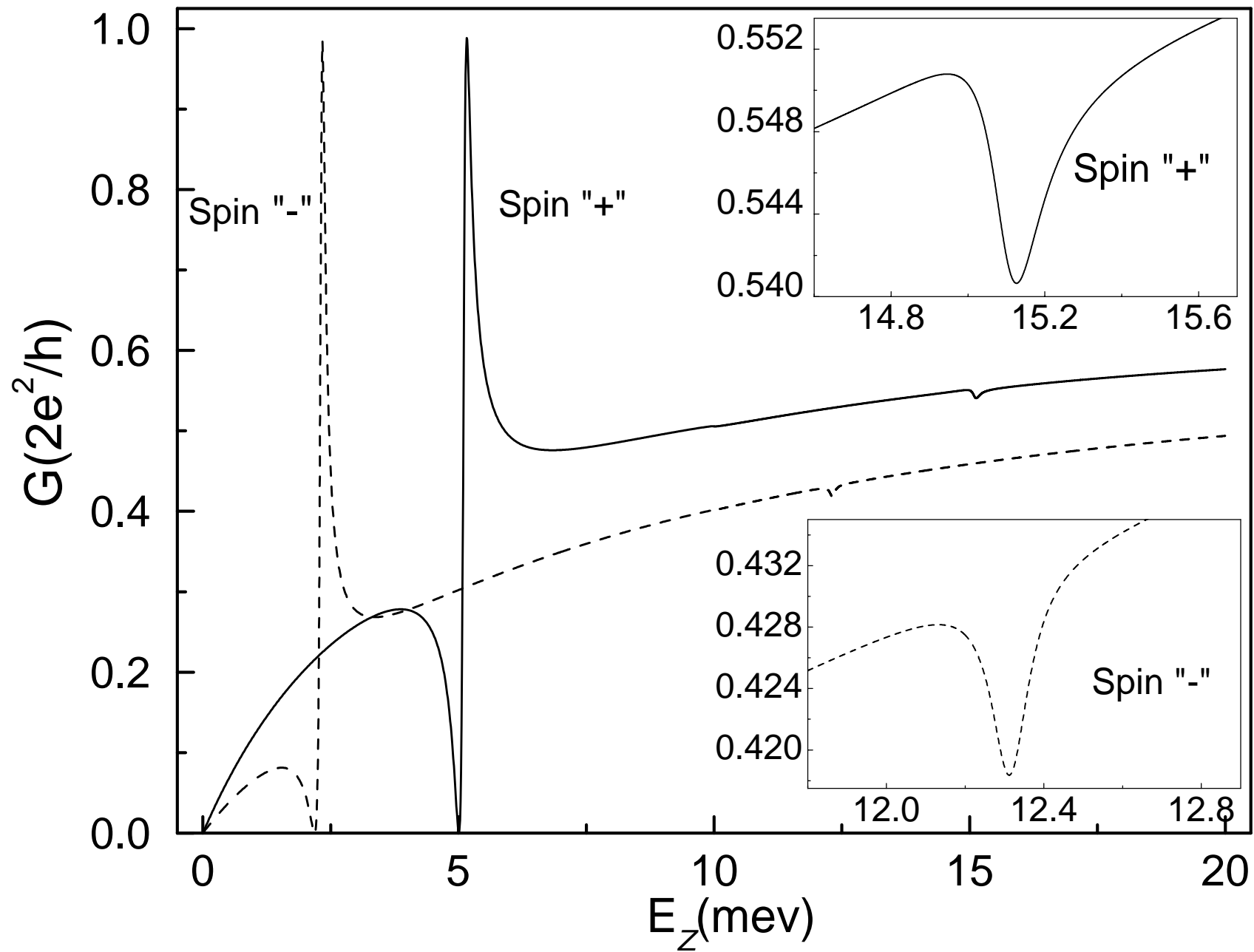
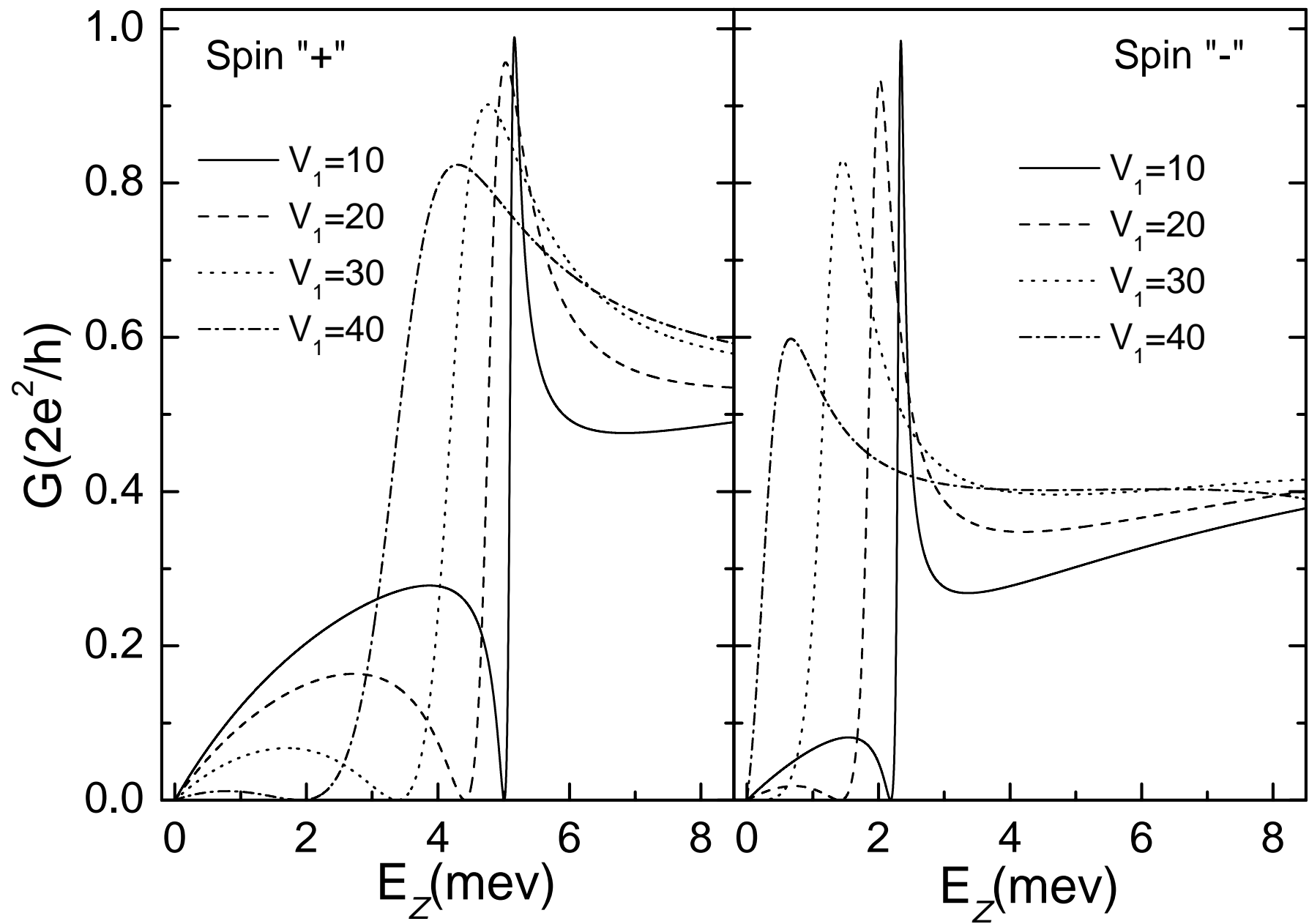
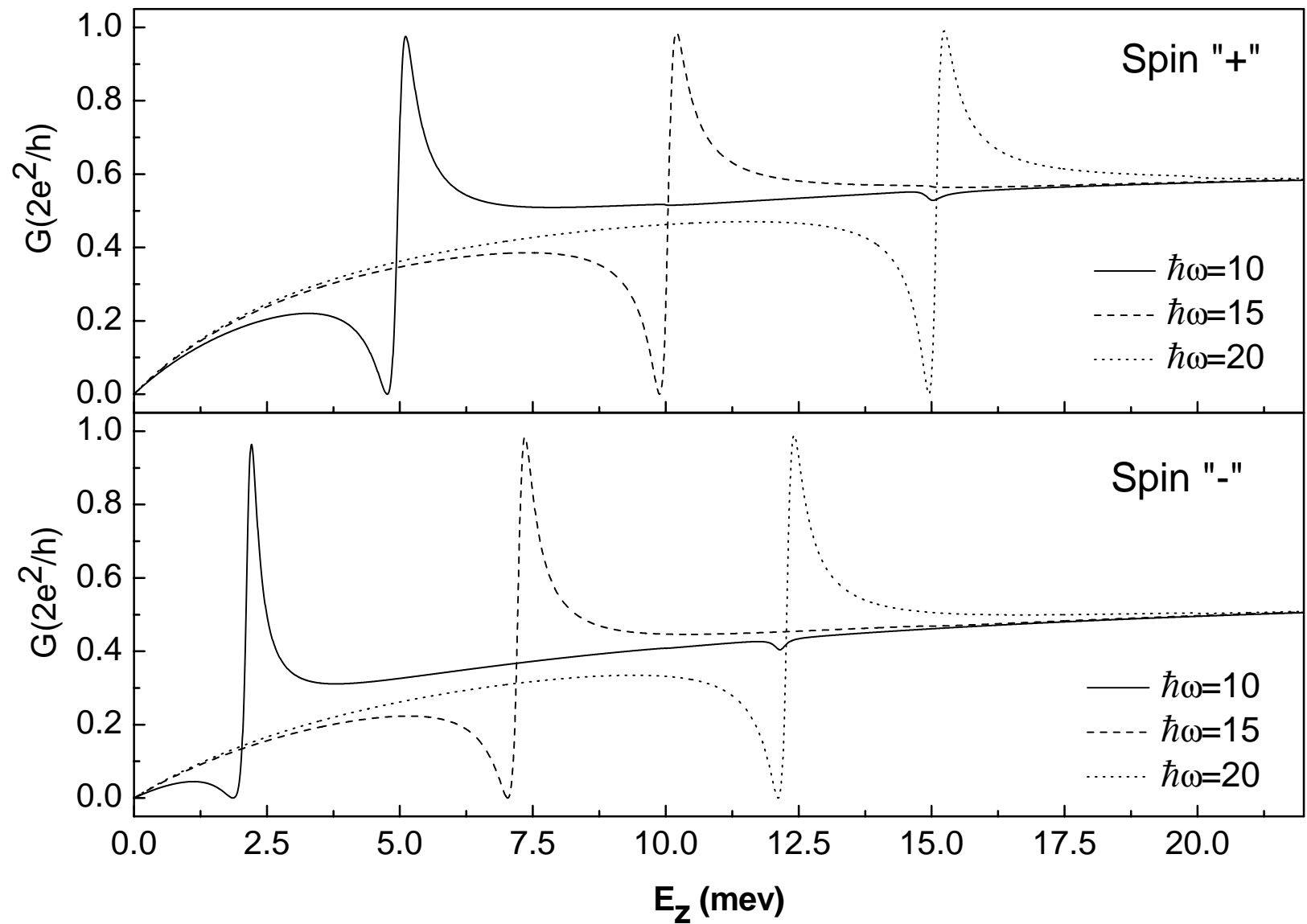
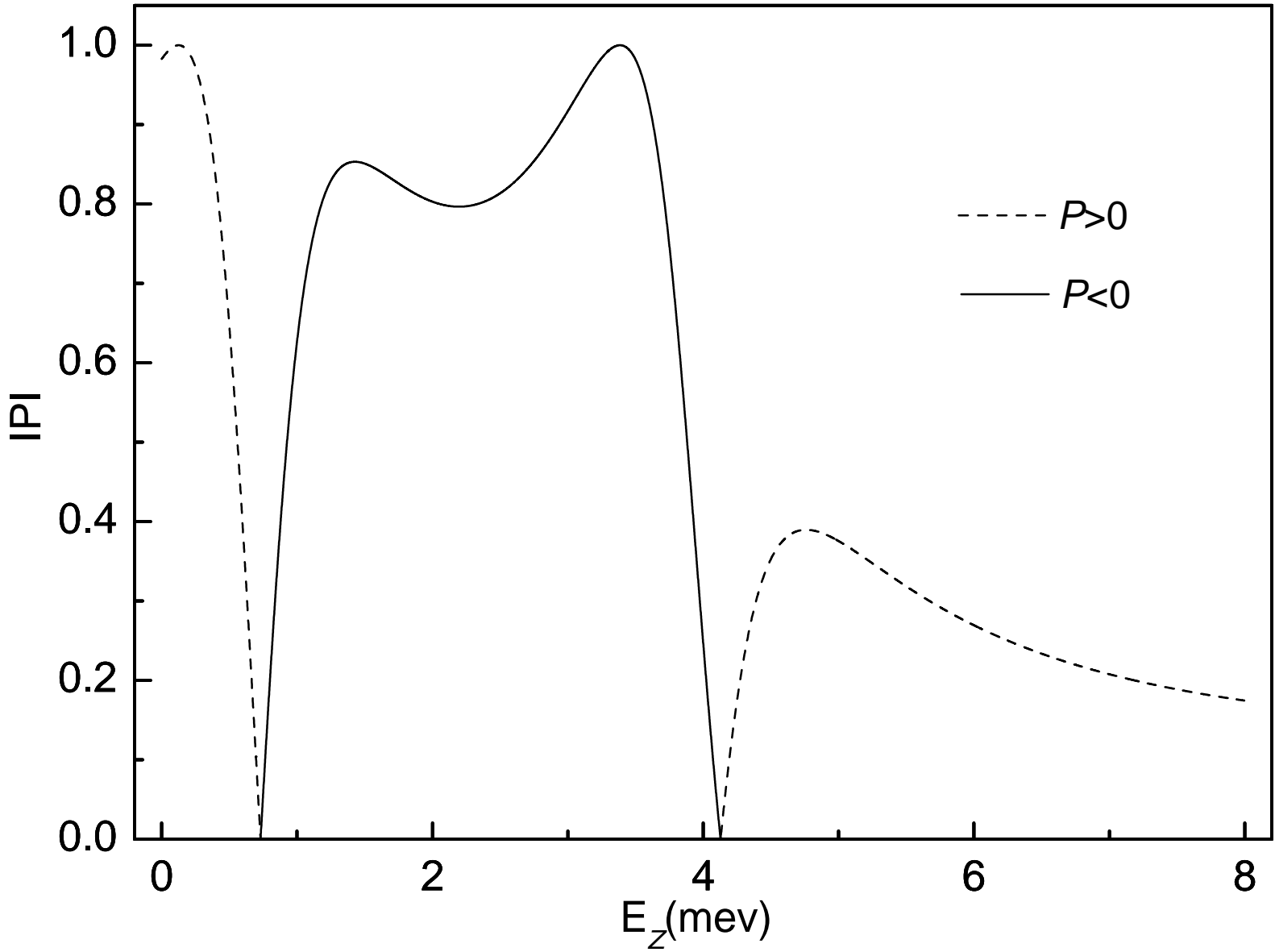
\end{document}